\documentclass[5p,times,twocolumn]{elsarticle}

\usepackage{amsmath,amssymb,graphicx,bm}
\usepackage{hyperref}
\usepackage{xcolor}

\journal{Physics Letters B}

\begin{document}

\begin{frontmatter}

\title{Structural Limitations on Constraining the Time Evolution of Dark Energy: \\ An Exact Linear Response Approach}

\author{Seokcheon Lee}
\affiliation{
Department of Physics, Sungkyunkwan University,\\
Suwon 16419, Republic of Korea
}
\ead{skylee@skku.edu}

\begin{abstract}
Cosmological constraints on a time-varying dark energy equation of state are fundamentally limited by the integral structure through which the equation of state enters cosmological observables. We rigorously derive the linear response kernel that maps perturbations in the equation of state $\omega(z)$ to comoving distance fluctuations $\delta D(z)$. By adopting a Fourier mode expansion $\delta \omega(z) = \sin(kz)$, we obtain the exact analytic form of the distance response in terms of Sine and Cosine integrals. We show that this mapping involves a double integration over redshift, which acts as an intrinsic low-pass filter with a characteristic $\sim k^{-2}$ scaling in redshift space. This structural limitation is visualized in a schematic diagram and confirmed by observational verification using the full covariance matrix of the Pantheon+ supernova dataset. Our analysis reveals a steep hierarchy in Fisher eigenvalues where the information content drops by an order of magnitude already at the second eigenmode. Consequently, distance-based probes effectively constrain only a single dominant mode of $\omega(z)$. This implies that the difficulty in constraining dynamical parameters such as $w_a$ is not due to data precision, but is a necessary consequence of the observable's integral nature, which renders it structurally blind to the instantaneous rate of change $d\omega/da$.
\end{abstract}


\end{frontmatter}

\section{Introduction}
\label{sec:intro}

Whether the dark energy equation of state evolves with cosmic time remains a central question in modern cosmology~\cite{Peebles:2002gy,Copeland:2006wr}. To address this, distance-based observables---including Type~Ia supernova luminosity distances~\cite{SupernovaSearchTeam:1998fmf,SupernovaCosmologyProject:1998vns,DES:2024jxu}, baryon acoustic oscillations~\cite{eBOSS:2020yzd,DESI:2024mwx,DESI:2025zgx,DESI:2025fii}, and cosmic microwave background~\cite{Planck:2018vyg} measurements---have been widely used to constrain the evolution of $\omega(z)$. These efforts often rely on phenomenological parametrizations, most notably the Chevallier-Polarski-Linder (CPL) ansatz \cite{Chevallier:2000qy, Linder:2002et}, or attempt model-independent reconstructions~\cite{Holsclaw:2010sk,Seikel:2012uu,Zhao:2017cud,Li:2025ops}. Despite the steadily improving precision of these data sets, constraints on the detailed time dependence of $\omega(z)$ remain weak and highly degenerate~\cite{Colgain:2024xqj,Sakr:2025daj,Abreu:2025zng,Gialamas:2025pwv,Lee:2025axp}.

This difficulty is not merely a consequence of limited data quality or specific parametrization choices. Distance probes access the dark energy sector indirectly through the cosmic expansion history, constraining $\omega(z)$ only after successive integrations over redshift~\cite{DESI:2024aqx,Zhang:2025bmk}. Early work emphasized that reconstructing $\omega(z)$ from distance measurements constitutes an ill-posed inverse problem, where small observational uncertainties are amplified during reconstruction \cite{Huterer:2000mj}. Direct reconstruction methods involving differentiation of noisy data face similar stability challenges \cite{Clarkson:2010bm}. Furthermore, analyses have demonstrated that luminosity-distance observables inherently average over the expansion history, making them insensitive to localized or rapidly varying behavior in $\omega(z)$ \cite{Maor:2000jy,Bassett:2004wz}.

To quantify these limitations, statistical techniques such as Principal Component Analysis (PCA) and Karhunen-Lo\`{e}ve eigenvalue decompositions have been employed \cite{Tegmark:1996bz, Huterer:2002hy}. These studies consistently show that distance data constrain only a small number of effective combinations of $\omega(z)$, leaving higher-order modes essentially unconstrained \cite{dePutter:2007kf,Crittenden:2011aa}.

A steep hierarchy in the information content of luminosity distance measurements has been noted previously. In particular, an early analysis by Astier~\cite{Astier:2000as} computed the eigenvalues of the Fisher matrix for supernova distance data and observed a
rapid suppression of higher-order modes. The present work provides an analytic explanation of this behavior by identifying the double-integral structure of the distance--redshift relation as its physical origin.

In this letter, we demonstrate that these limitations arise from a single underlying origin: the physical response kernel that maps $\omega(z)$ onto cosmological distances involves a double integration over redshift. This structure induces an intrinsic low-pass filtering of the dark energy equation of state. We make this statement explicit by deriving the exact linear-response kernel and providing an analytic solution for oscillatory modes. We visualize this mechanism schematically and validate the theoretical limit using the full covariance matrix of the Pantheon+ supernova dataset \cite{Scolnic:2021amr, Brout:2022vxf}. Our results confirm that distance-based probes alone are structurally insufficient to fully characterize dynamical dark energy, highlighting the necessity of complementary probes such as direct expansion rate measurements \cite{Jimenez:2001gg} or the growth of structure \cite{Linder:2005in, Zhao:2011te, Lee:2025hjw} to break these fundamental degeneracies.

\section{Theoretical Framework}
\label{sec:theory}

We consider a flat Friedmann-Lema\^{i}tre-Robertson-Walker universe. The background expansion rate $H(z)$ for the late universe is determined by the sum of energy densities
\begin{equation}
    H^2(z) = \frac{8\pi G}{3} \sum_i \rho_i(z) = H_0^2 \left[ \Omega_{m0}(1+z)^3 + \Omega_{\rm DE}(z) \right].
\end{equation}
We study the linear response of the observables to perturbations in the dark energy equation of state around a cosmological constant background ($\omega(z) = -1$). Let $\omega(z) = -1 + \delta \omega(z)$, where $|\delta \omega(z)| \ll 1$.

The evolution of the dark energy density $\rho_{\rm DE}$ is governed by the continuity equation, $d\ln \rho_{\rm DE} / dz = 3(1+\omega)/(1+z)$. Perturbing this equation to first order yields
\begin{equation}
    \frac{d}{dz} \left( \frac{\delta \rho_{\rm DE}}{\rho_{\rm DE}} \right) = \frac{3 \delta \omega(z)}{1+z}.
\end{equation}
Integrating from $z=0$ to $z$, we obtain the density perturbation
\begin{equation}
    \frac{\delta \rho_{\rm DE}(z)}{\rho_{\rm DE}(z)} = 3 \int_0^z \frac{\delta \omega(z')}{1+z'} dz'.
    \label{eq:rho_pert}
\end{equation}
Variation of the Friedmann equation, $2H \delta H = (8\pi G/3) \delta \rho_{\rm DE}$, leads to the fractional perturbation in the expansion rate
\begin{equation}
    \frac{\delta H(z)}{H(z)} = \frac{1}{2} \frac{\delta \rho_{\rm DE}(z)}{\rho_{\rm crit}(z)} = \frac{1}{2} \Omega_{\rm DE}(z) \frac{\delta \rho_{\rm DE}(z)}{\rho_{\rm DE}(z)}.
    \label{eq:var_Friedmann}
\end{equation}
Substituting Eq.~\eqref{eq:rho_pert} into Eq.~\eqref{eq:var_Friedmann}, we express the Hubble perturbation explicitly as an integral over the equation of state history
\begin{equation}
    \frac{\delta H(z)}{H(z)} = \frac{3}{2} \Omega_{\rm DE}(z) \int_0^z \frac{\delta \omega(z')}{1+z'} dz'.
    \label{eq:H_pert}
\end{equation}

The comoving distance is defined as $D(z) = \int_0^z c \, dz' / H(z')$, where $c$ is the speed of light. Note that the fractional perturbation of the luminosity distance $D_L = (1+z)D$ is identical to that of the comoving distance, $\delta D_L/D_L = \delta D/D$. The perturbation $\delta D(z)$ induced by $\delta H$ is
\begin{equation}
    \delta D(z) = - \int_0^z \frac{c \, dz_1}{H(z_1)} \frac{\delta H(z_1)}{H(z_1)}.
\end{equation}
Inserting Eq.~\eqref{eq:H_pert} into this expression yields a double integral structure
\begin{equation}
    \delta D(z) = - \int_0^z dz_1 \frac{c}{H(z_1)} \left[ \frac{3}{2} \Omega_{\rm DE}(z_1) \int_0^{z_1} dz' \frac{\delta \omega(z')}{1+z'} \right].
\end{equation}
The integration domain is the triangle defined by $0 \le z' \le z_1 \le z$. To isolate $\delta \omega(z')$, we exchange the order of integration (Fubini's theorem), changing the limits to $\int_0^z dz' \int_{z'}^z dz_1$. This leads to the linear convolution form
\begin{equation}
    \delta D(z) = \int_0^z K_D(z; z') \delta \omega(z') dz',
\end{equation}
where the kernel $K_D(z; z')$ encapsulates the background cosmology's geometric response
\begin{equation}
    K_D(z; z') = -\frac{3}{2} \frac{1}{1+z'} \int_{z'}^z \frac{c \, dz_1}{H(z_1)} \Omega_{\rm DE}(z_1).
    \label{eq:kernel}
\end{equation}
This kernel describes exactly how a localized perturbation $\delta \omega(z')$ at redshift $z'$ propagates to the observable distance at redshift $z$.

\begin{figure}[t]
\centering
\includegraphics[width=\columnwidth]{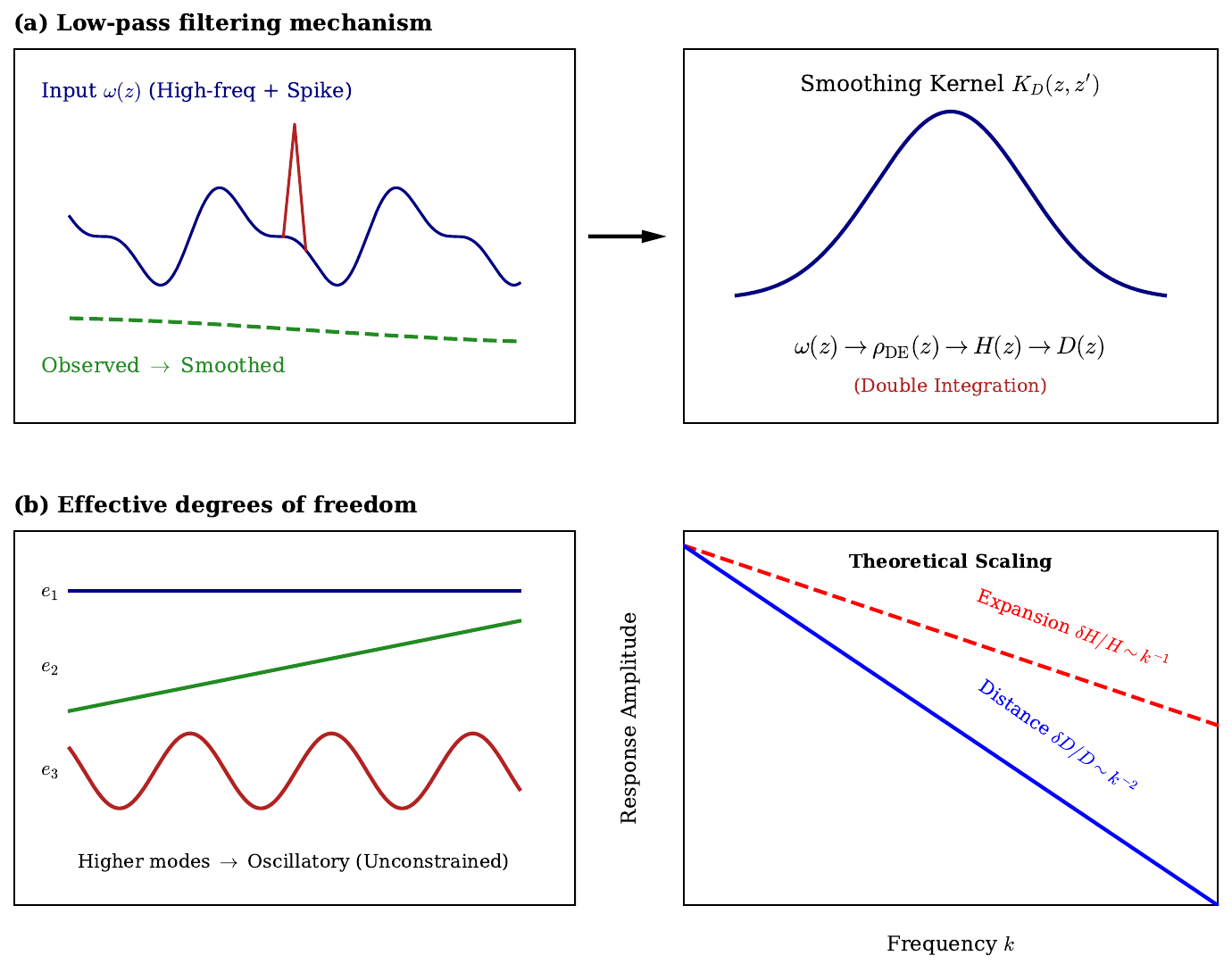}
\caption{
Schematic illustration of the structural low-pass filtering mechanism.
(a) High-frequency variations or localized spikes in the equation of state $\omega(z)$ (navy/red) are smoothed out by the broad double-integration kernel $K_D(z, z')$, resulting in a featureless observed distance response (green dashed).
(b) Consequently, a PCA reveals that only the first few smooth eigenmodes ($e_1, e_2$) are constrained.
The bottom-right panel illustrates the theoretical scaling of the response amplitude for expansion rate ($\delta H/H \sim k^{-1}$, red dashed) versus distance ($\delta D/D \sim k^{-2}$, blue solid), showing the steeper suppression in distance observables.
}
\label{fig:schematic}
\end{figure}

Throughout this work, we assume a spatially flat FLRW background. Furthermore, the matter density parameter $\Omega_{m0}$ is held fixed in order to isolate the structural response of distance observables to perturbations in the dark-energy equation of state. This choice is motivated by early studies~\cite{Astier:2000as}, which demonstrated that the luminosity distance derivatives with respect to $\Omega_{m0}$ and dark energy parameters exhibit similar shapes, leading to severe degeneracies. Allowing $\Omega_{m0}$ to vary would introduce additional degeneracies that further degrade the constraints, reinforcing our conclusion regarding the limited resolving power of distance probes, without altering the fundamental integral structure of the kernel.

\section{Exact Analytic Solution for Fourier Modes}
\label{sec:analytic}

To quantify the information loss rigorously, we analyze the linear response to a high-frequency oscillatory perturbation. We consider a single Fourier mode in redshift space given by $\delta \omega(z) = \sin(k z)$.
The perturbation to the expansion rate, Eq.~\eqref{eq:H_pert}, involves the integral $I_H(z) = \int_0^z \frac{\sin(k z')}{1+z'} dz'$. By introducing a change of variable $u = k(1+z')$, this integral can be expressed exactly in terms of the Sine Integral, $\text{Si}(x) \equiv \int_0^x \frac{\sin t}{t} dt$, and the Cosine Integral, $\text{Ci}(x) \equiv -\int_x^\infty \frac{\cos t}{t} dt$. The exact analytic solution is
\begin{align}
    I_H(z) &= \cos(k) \left[ \text{Si}(k(1+z)) - \text{Si}(k) \right] \nonumber \\
           &- \sin(k) \left[ \text{Ci}(k(1+z)) - \text{Ci}(k) \right].
    \label{eq:exact_IH}
\end{align}
Consequently, the fractional perturbation to the Hubble expansion rate is given exactly by $\delta H(z)/H(z) = \frac{3}{2} \Omega_{\rm DE}(z) \mathcal{S}_k(z)$, where we defined the mode response function $\mathcal{S}_k(z) \equiv I_H(z)$.

The resulting perturbation to the comoving distance $\delta D(z)$ is obtained by integrating the expansion rate perturbation
\begin{equation}
    \delta D(z) = -\frac{3}{2} \int_0^z \frac{c \, dz_1}{H(z_1)} \Omega_{\rm DE}(z_1) \mathcal{S}_k(z_1).
    \label{eq:exact_dD}
\end{equation}
Eqs.~\eqref{eq:exact_IH} and \eqref{eq:exact_dD} represent the exact linear response of the observable to the equation of state mode, valid for any background cosmology.

To demonstrate the structural suppression for high-frequency modes ($k \gg 1$), we utilize the asymptotic expansions of the special functions: $\text{Si}(x) \sim \frac{\pi}{2} - \frac{\cos x}{x}$ and $\text{Ci}(x) \sim \frac{\sin x}{x}$. Substituting these into Eq.~\eqref{eq:exact_IH}, the leading-order behavior of the response function is derived as
\begin{equation}
    \mathcal{S}_k(z) = -\frac{1}{k} \left[ \frac{\cos(k(1+z))}{1+z} - \cos(k) \right] + \mathcal{O}(k^{-2}).
\end{equation}
This explicitly shows that the response of the expansion rate is suppressed by a factor of $k^{-1}$. Furthermore, when evaluating the distance perturbation $\delta D(z)$, the oscillatory term $\cos(k(1+z))$ in the integrand undergoes a second integration. Through integration by parts, $\int \cos(kz) dz \sim k^{-1} \sin(kz)$, this introduces an additional suppression factor of $1/k$. Thus, the amplitude of the distance perturbation scales as
\begin{equation}
    \delta D(z) \sim \mathcal{O}(k^{-2}).
\end{equation}
This mathematically confirms that the double-integral mapping from $\omega(z)$ to $D(z)$ acts as a second-order low-pass filter. The exact analytic forms provided here prove that rapid temporal variations in $\omega(z)$ are structurally suppressed by the square of their frequency, rendering them observationally indistinguishable from noise regardless of data precision.

\section{Observational Verification with Pantheon+}
\label{sec:Obs}

To verify that the derived theoretical limitations persist in high-precision observational data, I performed a Fisher matrix analysis using the full covariance matrix of the Pantheon+ Type Ia supernova dataset ($N=1701$) \cite{Scolnic:2021amr, Brout:2022vxf}.

The Pantheon+ dataset was selected for this analysis to explicitly isolate structural limitations from issues related to data sparsity. Unlike baryon acoustic oscillation data, which are often compressed into discrete redshift bins, Pantheon+ provides individual data points with dense sampling, particularly in the low-redshift regime. This allows for a robust test of the integral kernel: if the weak constraints on dynamical dark energy were due to insufficient data points or coarse binning, the high number density of the Pantheon+ sample should theoretically recover higher-order modes. By utilizing the unbinned covariance matrix, this analysis demonstrates that the loss of information is a fundamental consequence of the distance-redshift mapping rather than a statistical artifact.

For the numerical analysis, the perturbation $\delta \omega(z)$ is expanded using a Chebyshev polynomial basis of order 15. While the Fourier basis was employed in Sec.~\ref{sec:analytic} for its analytical clarity, the Chebyshev basis is adopted here for numerical stability. Chebyshev polynomials form a complete and orthogonal basis on a finite interval ($0 \le z \le z_{\rm max}$), which minimizes boundary artifacts such as the Gibbs phenomenon when reconstructing non-periodic functions \cite{dePutter:2007kf, Crittenden:2011aa}.

The Fisher information matrix is related to the parameter covariance matrix by $\mathbf{F} = \mathbf{C}^{-1}$. Thus, the eigenvalues $\lambda_k$ of the Fisher matrix correspond to the inverse variance of each eigenmode ($\lambda_k = 1/\sigma_k^2$). A rapid decay in $\lambda_k$ indicates an exponential increase in the uncertainty of higher-order modes.

\begin{figure}[t]
\centering
\includegraphics[width=\columnwidth]{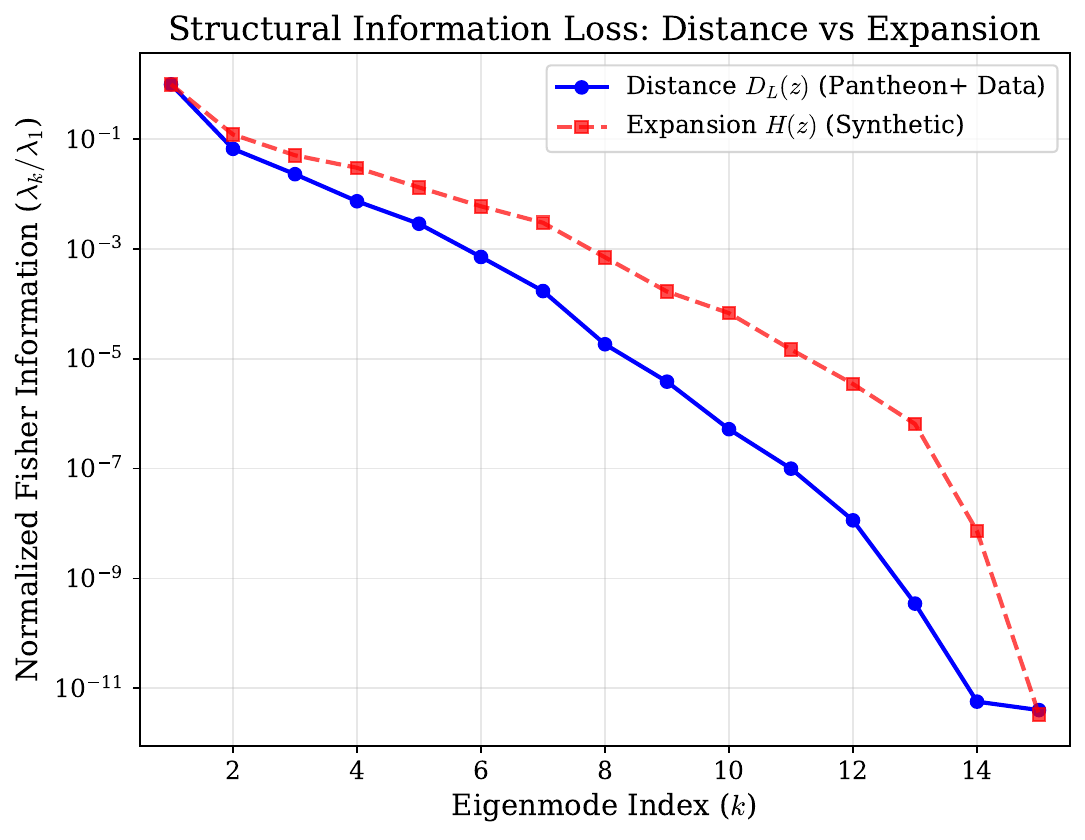}
\caption{
Normalized Fisher information eigenvalues for the reconstruction of $\omega(z)$. The blue solid line shows constraints from Pantheon+ luminosity distances ($N=1701$), while the red dashed line represents a synthetic expansion rate ($H(z)$) dataset with identical redshift distribution. The steep decay in the blue curve confirms the severe information loss induced by the double integration $\omega(z) \to D_L(z)$. While $H(z)$ measurements show a $k^{-1}$ suppression, distance measurements exhibit a stronger $k^{-2}$ suppression, leaving only the first eigenmode well-constrained.
}
\label{fig:eigenvalues}
\end{figure}

Fig.~\ref{fig:eigenvalues} presents the eigenvalue spectrum derived from the Pantheon+ data (blue solid line). The spectrum exhibits a steep decay consistent with the predicted $\sim k^{-2}$ scaling, revealing a distinct hierarchy:
\begin{itemize}
    \item The first eigenvalue $\lambda_1$ (normalized to 1) corresponds to the single dominant mode constrained by the data, effectively representing the pivot equation of state $\omega_p$~\cite{Albrecht:2006um}.
    \item The second eigenvalue drops to $\lambda_2 / \lambda_1 \approx 0.08$, implying that the constraint on the second mode is weaker by an order of magnitude compared to the constant component.
    \item For the third mode, the information content falls to $\lambda_3 / \lambda_1 \approx 0.02$, rendering it effectively unconstrained.
\end{itemize}

For comparison, a synthetic dataset for the expansion rate $H(z)$ (red dashed line) was constructed with the same redshift distribution and fractional precision as the Pantheon+ sample (\textit{i.e.}, $\sigma_H(z_i)/H(z_i) \simeq \sigma_{D_L}(z_i)/D_{L}(z_i)$). This ensures that the comparison reflects the structural difference between the single-integral ($H$) and double-integral ($D_L$) kernels, rather than differences in data volume. Although $H(z)$ retains slightly more information ($\lambda_2/\lambda_1 \approx 0.15$) due to its lower integration order, it still displays characteristic low-pass behavior. 

It is also important to distinguish between the number of basis functions ($N=15$) and the number of \textit{effective} degrees of freedom.  The observed hierarchy $\lambda_1 \gg \lambda_2 \gg \lambda_3$ therefore implies
that distance probes constrain only a single effective degree of freedom of the
dark-energy equation of state.

The same qualitative behavior is also observed in DESI DR2 BAO data. Analyses of DESI DR2 measurements demonstrate that introducing additional  degrees of freedom in the dark energy equation of state beyond the first two modes yields no significant improvement in goodness-of-fit (see Fig.~6 of Ref.~\cite{DESI:2025fii}). This saturation explicitly confirms that the pronounced eigenmode hierarchy found here is not specific to supernova distances, but reflects a structural limitation of late-time cosmological 
observables. 

This limitation is not specific to parametrizations like CPL ($\omega_0, \omega_a$), but represents a fundamental inability of distance data to constrain time-varying dark energy ($\omega \neq \text{const}$). Even if the equation of state undergoes rapid temporal variations, the observable distance reflects only the integrated history, smoothing out such features. Consequently, constraining dynamical degrees of freedom using distance data alone is equivalent to attempting to extract information that has been structurally suppressed by the integration kernel.

\section{Conclusion}
\label{sec:con}

We have presented an exact derivation of the linear response of cosmological distances to perturbations in the dark energy equation of state. By solving the convolution integrals analytically, we proved that the distance observable acts as an intrinsic low-pass filter, suppressing high-frequency information with a characteristic $k^{-2}$ scaling. This theoretical prediction is robustly supported by our numerical analysis of the Pantheon+ dataset, which exhibits a rapid decay in Fisher information eigenvalues.

The resulting hierarchy, $\lambda_1 \gg \lambda_2$, establishes that distance-based probes are structurally capable of constraining only one effective degree of freedom of $\omega(z)$. It is crucial to emphasize that this result is not an artifact of the Principal Component Analysis (PCA) method or the specific basis chosen. Rather, PCA merely exposes the fundamental physical reality that distance observables, being integral in nature, are structurally blind to the instantaneous rate of change $d\omega/da$. Even if the equation of state undergoes rapid temporal variations, the observable distance reflects only the smoothed, integrated history of these changes. Consequently, the inability to constrain dynamical dark energy is a structural limitation of the probe itself, independent of data precision. Future efforts to confirm dynamical dark energy must therefore rely on combining distance measurements with observables that possess inherently different kernel structures, such as the growth of structure which depends on the derivative of the expansion history.

\section*{Acknowledgments}

This work is supported by the Basic Science Research Program through the National Research Foundation of Korea (NRF), funded by the Ministry of Science and ICT under Grant No.~NRF-RS-2021-NR059413 and NRF-2022R1A2C1005050.

\section*{Declaration of generative AI and AI-assisted technologies}
The author declares that no generative AI or AI-assisted technologies were used
in the writing of this manuscript.

\section*{Conflict of interest}
The author declares no competing interests.

\bibliographystyle{elsarticle-num}
\bibliography{references}

\end{document}